*Original Article*

# Sentiment Analysis of COVID-19 Public Activity Restriction (PPKM) Impact using BERT Method

Fransiscus[1], Abba Suganda Girsang[2]

[1,2] *Computer Science Department, BINUS Graduate Program – Master of Computer Science,*
*Bina Nusantara University, Jakarta, Indonesia.*

[1]*Corresponding Author : fransiscus002@binus.ac.id*

***Abstract*** *- Covid-19 has grown rapidly in all parts of the world and is considered an international disaster because of its wide-reaching impact. The impact of Covid-19 has spread to Indonesia, especially in the slowdown in economic growth. This was influenced by the implementation of Community Activity Restrictions (PPKM) which limited community economic activities. This study analyzes the mapping of public sentiment towards PPKM policies in Indonesia during the pandemic based on Twitter data. Knowing the mapping of public sentiment regarding PPKM is expected to help stakeholders in the policy evaluation process for each region. The method used is BERT with IndoBERT specific model. The results showed the evaluation value of the IndoBERT f-1 score reached 84%, precision 86%, and recall 84%. Meanwhile, f-1 scores 70%, 72% precision, and 70% recall for evaluating the use of SVM. Multinominal Naïve Bayes evaluation shows an f-1 score of 83%, precision of 78%, and recall of 80%. In conclusion, the BERT method with the IndoBERT model is proven to be higher than classical methods such as SVM and Multinominal Naïve Bayes.*

## 1. Introduction

Based on research by Murad et al. 2020, the Covid-19 pandemic has grown rapidly in all parts of the world and is considered an international disaster because of its wide-reaching impact. All lines of life are affected by the spread of this virus in many countries. Various measures have been taken to reduce the impact on people's health, socio-economic, education, and habits. Indonesia's economy throughout the year slowed down to minus 5.3 percent in the second quarter of 2020, and the average growth reached minus 2.1 percent in 2020 [2].

The implementation of PPKM will be considered effective if viewed through the perspective of health analysis. Still, their understanding will be contradictory if viewed from an economic perspective experienced by the community. The urgency of PPKM implementation must be well conveyed to the public to support the percentage of successful PPKM implementation, by setting sanctions for people who violate PPKM can help smooth the implementation of PPKM policies. One of the sectors most affected by the PPKM policy is micro-business in society.

Based on survey data on 206 small and medium entrepreneurs (UMKM) in Greater Jakarta, as many as 82.9% of UMKM entrepreneurs experienced a negative impact on the pandemic conditions. The impact of a 30% decrease in turnover was also experienced by 63.9% of the affected UMKM [3]. The rapid growth of the positive number of the Covid-19 virus and the policy of restrictions from the government is a challenge for UMKM actors. Of course, this can potentially produce a chain economic impact in the future.

Moreover, implementing government policies related to pandemics must be in line with the conditions of the people in various affected areas. The implementation of a unilateral policy will certainly harm many parties. A Survey is needed to see how the public responds to this policy, responses, reactions, or opinions from the public.

In addition to survey data, people's opinions and reactions share their views on various social media sites such as Twitter [4]. Research related to Twitter sentiment on government policies shows that by knowing public sentiment, the government can take quick action to re-evaluate the policies taken, especially on topics with a high negative rate based on opinions on Twitter [5].

A huge number of Twitter users will certainly generate many opinions related to many things every day, including responses related to PPKM. Therefore, in this study, the authors used Twitter data as a source. To get the data, the author uses data mining techniques [6]. According to Gupta and Chandra 2020, data mining is an efficient extraction technique used to analyze qualitative data on a large scale. It can be done repeatedly for the benefit of solving a problem. This research uses data mining to extract public Twitter information to analyze public opinion and sentiment on PPKM policies in various regions. This opinion is then processed into an expression using the sentiment analysis method.

According to Diaz-Garcia, Ruiz, and Martin-Bautista 2020, sentiment analysis includes text mining techniques, natural language processing, and automatic learning that focuses on obtaining sentimental aspects from the text. The sentiment result can be obtained using machine learning methods [9]. Machine learning can be broadly defined as a computational method that uses experience to improve performance or make accurate predictions. The experience referred to can be in the form of existing data (training set) or the process of system interaction with the environment [10].

In this study, the method used is Bidirectional Encoder Representations from Transformers (BERT). BERT is a machine learning model created to improve accuracy in Natural Language Processing (NLP). BERT was developed by Google AI Language researchers in 2018. This method was developed by collaborating deep learning techniques with methods such as UMLFiT, OpenAI Transformers, and Transformers [11]. BERT is divided into two models such as BERTBASE (12-layer encoder, 12 heads attention, hidden size 768, and 110M parameters) and BERTLARGE (24 layers, 16 heads attention, hidden size 1024, and 340M parameters) [12].

Previous research related to sentiment analysis on social media in China shows the accuracy of the BERT model (75.65%) outperforms classical algorithms such as SVM (70.66%), Naïve Bayes (66.97%), Logistic Regression (70.02%), CNN (71.19%), and LSTM (57.73%). BERT is effectively used in sentiment analysis because this algorithm continues to be developed based on NLP, allowing it to be applied in processing public opinion with big data [13].

Previous study regarding identifying sentiments from opinions related to Covid-19 vaccination based on Twitter data. This study uses TF-IDF as a feature extraction method and compares 6 classification algorithms, namely Naive Bayes, Random Forest, SVM, Bi-LSTM, BERT, and CNN. The performance measurement results show that the BERT algorithm has the highest accuracy of 78.94%, and the lowest performance is demonstrated by CNN, with an accuracy of 69.01%. The classification results show very high neutral sentiment reaching 70% of the data, followed by positive sentiment at 20%, then negative sentiment at 10% [14].

Chandra and Saini 2021 analyzed sentiment on Twitter data regarding presidential candidates Biden and Trump in the 2020 election in the United States. This study uses the LSTM and BERT methods to classify public sentiment towards the two figures. The data used is quite large, namely 1.1 million tweets with geolocation. The study results show that Biden's electability surpasses Trump, with performance measurements showing that BERT is the best model, with an accuracy of 87.45% and an F1 score of 75.7%. In comparison, LSTM has an accuracy of 85.62% F1 score of 68.6%.

Previous research by Nugroho et al. 2021 [16] related to sentiment analysis of user reviews of mobile-based applications on Google Play using fine-tuned IndoBERT. In this study, a comparison was made of the effectiveness of the data labelling process on fine-tuning BERT such as BERT-M and IndoBERT with traditional machine learning models such as kNN, SVM, Naïve Bayes, Decision Tree, and Random Forest. Two experiments were carried out with both lexicon-based and scoring-based labeling on the training data. The result is that the BERT method is superior, especially IndoBERT with lexicon-based training data, which has the highest accuracy of 84%.

Previous research involving the BERT method shows that BERT is the best method with the best accuracy. However, the accuracy is still not optimal because it still depends on other supporting processes, such as data quality on BERT training processes. However, when viewed from the results and process, BERT tends to be more thorough, especially in understanding the context and relevance of a sentence. Therefore, the author chooses BERT as the suitable method and is still very open to innovation and collaboration to increase accuracy.

In this study, the author will use the IndoBERT transformer model as a reference. IndoBERT is a comprehensive Indonesian language benchmark consisting of seven assignments for the Indonesian language. These benchmarks are categorized into three pillars of NLP tasks, namely morpho-syntax, semantics, and discourse. IndoBERT works based on rich data where more than 200 million words are trained [17]. Previous research by Nugroho et al. 2021 [16] shows high accuracy for IndoBERT with lexicon-based training data preparation. However, lexicon-based training data need more validation. In this study, the author adds a manual validation process to verify that all labelled training data is already accurate to build the final model.

### 2.1. Data Collection

Based on figure 1, it is known that the research stages begin with data collection. As a classification-based application, it requires a data source. The author uses the Twitter dev API for the ad-hoc or stream data retrieval process using Python. This data retrieval process can be done with a lag time of 15 minutes and a maximum of up to 900 requests for data. Keywords used in this study 'PPKM' and 'Jakarta' with a total of up to 50,000 records starting from January 1, 2021, until March 30, 2022.

## 2. Materials and Methods

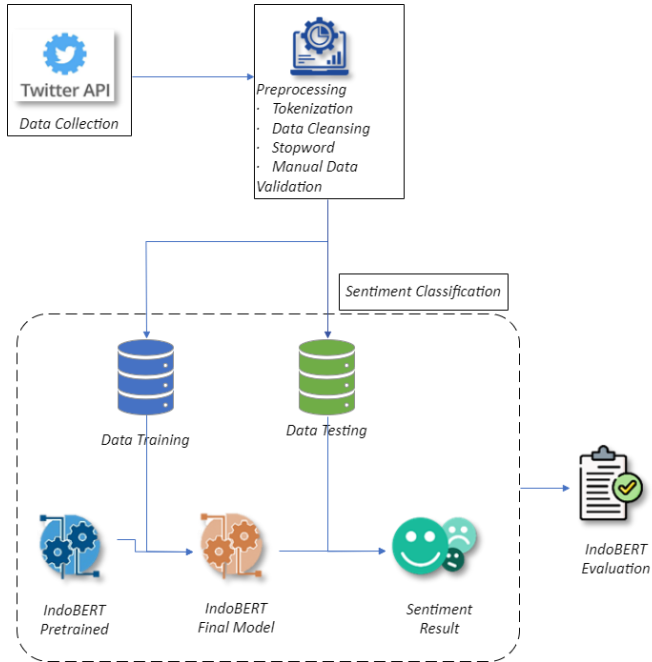

**Fig. 1 Research Process**

### 2.2. Data Pre-processing
At this stage, the author makes a preprocessing process to clean and normalize the input data to facilitate the classification process. The preprocessing process consists of several steps in it.

#### 2.2.1 Tokenization
The purpose of this tokenization is to break a sentence into pieces of words [18]. Tokenization is useful in order to extract meaning from a text. For example, in a sentence, authors want to detect nouns and verbs in the sentence, or authors want to find the name of a person whose dimensions are in a sentence.

#### 2.2.2 Data Cleansing
Data cleansing is the process of cleaning data from noise which aims to facilitate the labelling and classification process on Twitter data. Data Cleansing consists of several stages of data cleaning by applying Regular Expression (RegEx), which aims to simplify the data classification process.

#### 2.2.3. Stopword Removal
Stopword removal is the process of removing noise in the form of hashtags, symbols, URLs, content, and words that have no clear meaning. The elimination of meaningless words is done based on the dictionary in the program [19]. In this study, stopwords were done by deleting meaningless words based on the existing dictionary on the http://ranks.nl/stopwords/ site.

#### 2.2.4 Manual Validation
After the data has been successfully cleaned at the data cleansing and stopword stages, the author performs manual validation to ensure that the initial data follows the PPKM context using Microsoft Excel. At this stage, two validations are necessary, such as data duplication and PPKM correlation. Duplicate data will be eliminated to ensure the tweet's unique value should be related to PPKM. In a special case, some tweets mention the word PPKM, but the whole tweet doesn't talk about PPKM. This case should be eliminated to avoid invalid training data.

### 2.3. Classification Process
#### 2.3.1 Data Training Initiation
After the cleansing process is completed, the next step is training data initiation. The initial classification uses the lexicon method with R programming. This lexicon is very simple by calculating the weight of positive, negative, and neutral words from tweets. Table 1 shows examples of positive and negative words collected by the author from http://ranks.nl/stopwords/indonesian.

After determining the dictionary of positive and negative words, the author builds a simple classification model with R programming. This model works by comparing positive and negative word searches in tweets. Every time a negative word is found, the sentiment score is -1; for every positive word, the sentiment score is +1. Furthermore, this value will be accumulated, and the calculation of the final value will be carried out. A final score of more than 0 is classified as positive, while a final score of less than 0 is classified as negative, and a value equal to 0 is classified as neutral. Sentiment classification results by lexicon are not entirely accurate. Therefore, the author again conducted a manual review to ensure the sentiment results were correct. This process is essential to ensure the accuracy of the prediction model.

**Table 1. Positive and negative words corpus**

| Positive word | Negative word |
|---|---|
| bersuka cita | barbar |
| bersuka ria | basi |
| bersyukur | bau |
| dapat diandalkan | bebal |
| dapat dipercaya | beban |
| dapat diraih | bejat |
| dapat disesuaikan | bekas |
| daya tarik | bekas luka |
| dengan mewah | nepotisme |
| dengan senang hati | neraka |
| dengan sopan | neurotik |
| dermawan | ngambek |

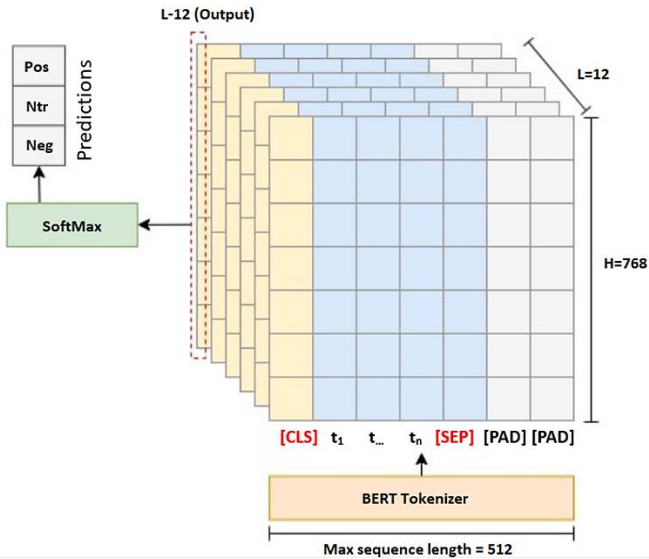

Fig. 2 BERT Fine-Tuning process

*2.3.2.. Model Initiation*

The data resulting from the previous stage's lexicon-based classification became the IndoBERT model's input. Figure 2 shows BERT fine-tuning process according to Nugroho et al., 2021[2].

The training process is not carried out in the early stages, but the training process starts by using a model that has been previously trained. In this study, the authors used the IndoBERT-base p1 reference Hugging Face model as the initial model. Then this model will be retrained for optimization in a new task, namely sentiment classification with PPKM context. This retraining process is referred to as fine-tuning. In this process, the BERT model will receive a sequence of words or sentences as input which will be processed in the encoder stack. Each encoder applies self-attention and provides output through a feed-forward network which the next encoder will follow. This process will continue 12 times according to the IndoBERT-base model.

In the next stage, after passing through all encoders, each token in each position will provide an output in the form of a vector with a size of 768. In the sentiment analysis task, the output given to the first token position is [CLS]. At the same time, a [SEP] token must be added at the end of the sentence [16]. The last layer in the classifier layer produces logits. Logits are output in the form of rough probability predictions of the sentences to be classified. Next, softmax will convert the logits into probabilities by taking the exponents of each logit value so that the total probability is exactly 1. This fine-tuning process can be done by adjusting the hyperparameters. In the BERT training process, the author uses 32 batch sizes, 10 epochs, and a learning rate (Adam) 3e-6.

*2.4. Testing & Visualization*

The next stage is to try the model with input data. In this study, the input data is in the form of tweets. Based on this process, an entity identification process will be generated where in this study, the author will focus on the location entity. Of course, the process of introducing this entity will depend on the quality of the training and previous training data. Therefore, in this study, the authors will continue improving the data quality and training process to produce better accuracy.

Figure 3 shows the classification of sample flow data using a pretrained IndoBERT model. Before fine-tuning, the sample tweet will be tokenized using the IndoBERT input format. Input formatting required a tokenizer by adding a special token for each sentence. A [CLS] token must be added at the beginning of each sentence for classification. At the same time, a [SEP] token must be added at the end of the sentence [16]. Tokenized sentences will be processed by a transformer block that includes an encoder. The transformer encoder learns and stores the tweet's semantic relationship and grammatical structure information, and the Lexicon-based class is based on this input [20]. Finally, softmax classifies the tweet as a sentiment result based on the information. The sentence "Jualan saya rugi selama PPKM" was classified as a Negative class in this case.

The author visualizes the data by making N-grams of the uni-gram type and word cloud. N-grams are combinations of several words used together in the text. A unigram has N=1, and the rationale behind n-grams is that language structure [21]. A word cloud is essentially a visual representation of text. Word clouds are used in many ways. It can usually be done with pure text compression. In this case, the word cloud represents words that occur more frequently [22].

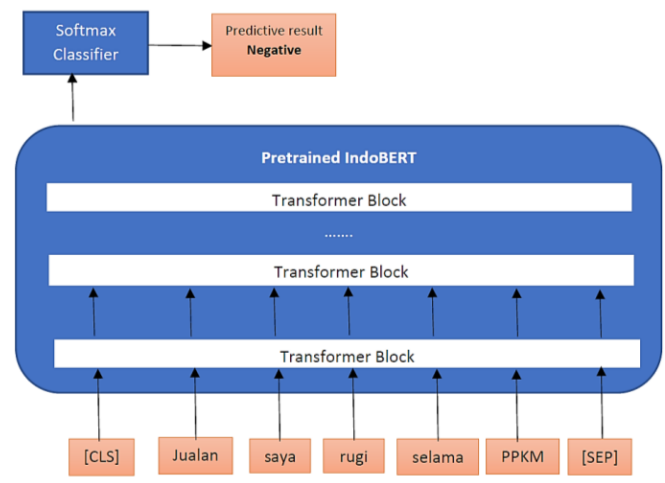

Fig. 3 Sentiment Analysis IndoBERT Example

Data visualization shows the most common issues in tweets about PPKM in the Jakarta area. With this issue, the author knows what opinions often arise according to their respective sentiment classes.

*2.5. Evaluation*

After the model is formed and the classification results are obtained, it is necessary to evaluate the model that has been formed. This evaluation aims to assess the algorithm's accuracy in the appointed case study to determine whether an algorithm can work well. Result performance measurement consists of a precision, recall, and F1-Score [23]. Precision means the ratio of samples flagged as positive to samples correctly flagged as positive. The recall is the ratio of the number of positively flagged samples to the total number of positively flagged samples. F-value is used as a scoring metric to analyze the sentiment classification view [24]. It is necessary to calculate the values to know the True Positive (TP), False Positive (FP), True Negative (TN), and False Negative (FN) values. They can be calculated as shown in Eq.(1), Eq.(2) and Eq.(3).

$$Precision, p = \frac{TP}{TP+FP} \quad (1)$$

$$Recall, r = \frac{TP}{TP+FN} \quad (2)$$

$$F - Score = \frac{2TP}{2TP+FP+FN} \quad (3)$$

The evaluation value should be compared with other methods, and whether BERT has superior performance compared to other classification methods that have been used in the same cases and data can be seen. In this stage, the authors compare the Naïve Bayes and SVM algorithms. In the initial research process, the authors tried to execute the Naïve Bayes and SVM algorithms, resulting in faster execution times than current methods such as Convolution Bi-directional Recurrent Neural Network (CBRNN) and Bi-directional Long Short-Term Memory (BiLSTM). As a comparison, the average training time required in Naïve Bayes and SVM training is under 15 minutes. Still, with the same data, CBRNN and BiLSTM need a much longer time, with the fastest time being 5 hours using the author's hardware and software. This execution constraint several times made the tools used by the author overloaded so that they had to repeat the test and made the research process less effective when applied to multilevel research such as sentiment mapping.

Several studies have shown that SVM and Naïve Bayes are still competitive in terms of performance compared to the latest methods, such as LSTM. Research by Nikmah et al. 2022 shows that the accuracy of SVM (86.54%) and Naïve Bayes (85.45%) can outperform LSTM (84.62%). In this study, the quality of accurate training data is decisive. Meanwhile, the authors used accurate training data in this study by carrying out a manual validation process for all data.

## 3. Results and Discussion

In this study, the author collects 50.000 raw data by Twitter API with the keyword 'PPKM' and 'Jakarta'. After the Twitter data is collected, it needs to be preprocessed. This stage consists of cleaning data, stopwords, and manual data validation. As explained in chapter 2, this process aims to clean and normalize the input data to simplify the classification process. In addition, the data collected still needs to be selected with several criteria that ensure the authors only use data related to PPKM. All data that does not fit into the criteria will be deleted manually, resulting in 5,315 data.

Before conducting the IndoBERT model training, the author will carry out the process of classifying data has been processed into 3 classes (positive, neutral, negative) with the Lexicon method, where the results of this model classification will be justified manually with the author to ensure that the classification results are precise and accurate. This is done to ensure that the data will produce maximum output. Based on this process, the authors manually justified sentiment classes in 5315 tweet data, resulting in 3,590 negative sentiments, 800 positive sentiments, and 925 neutral sentiments.

The author continues to train the model with the IndoBERT library as a classification algorithm. The author divides the 5315 data into 4877 training data, 293 validation data, and 145 testing data in the training process. The author uses 10 epochs to build a model in the training process. Figure 4 shows how the IndoBERT model of the accuracy of the BERT fine-tuning learning process continues to increase from epochs 1 to 10.

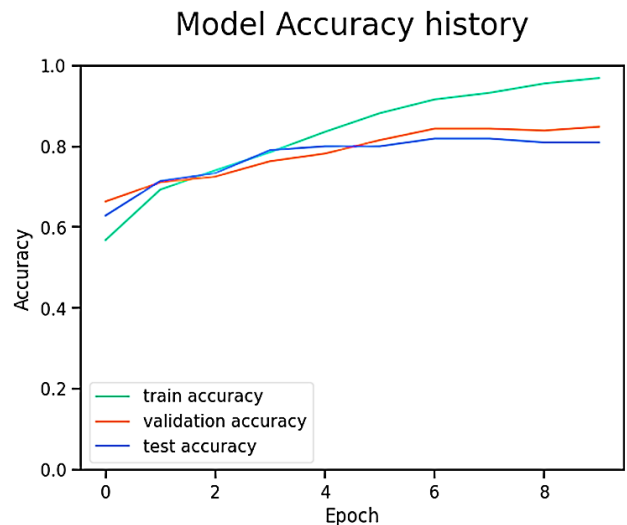

**Fig. 4 BERT Fine-Tuning Accuracy History**

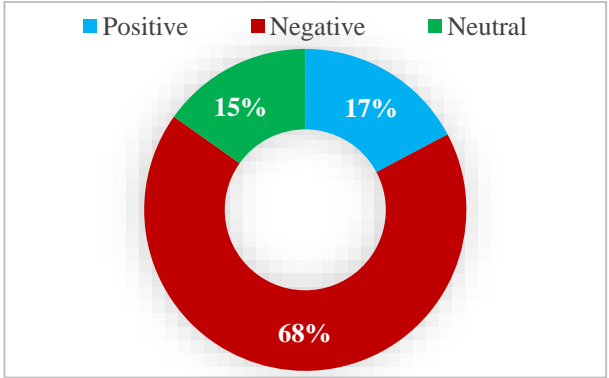

**Fig. 5 BERT Testing Sentiment Result**

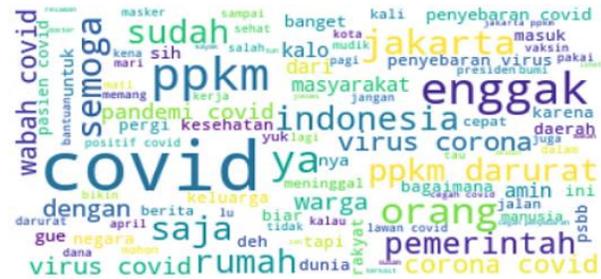

**Fig. 6 Word Cloud**

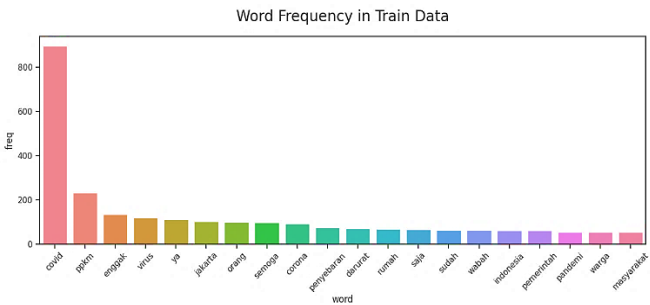

**Fig. 7 Word N-gram**

Testing and validation accuracy remains flat value due to there are several data negative and positive training. However, figure 5 shows the testing process is dominated by negative sentiment. This shows a negative society response to the PPKM policy in Jakarta based on tweet data. Figure 6 shows the word cloud that represents covid, ppkm, 'enggak' (means no) as a dominant value. The interesting word is "pemerintah" (means government) also mentioned. Figure 7 shows N-gram visualization, same as the word cloud, N-gram shows covid and ppkm on the top list. Word "pemerintah" also mentioned in this N-gram. By this visualization, authors can identify the most common issues in tweets about PPKM from the frequent word that appeared.

After defining the performance calculation, BERT needs to be compared with other methods with the same data. Compared with other methods, whether BERT performs better than other classification methods used in the same case and data can be seen. At this stage, the authors chose the Multinominal Naïve Bayes algorithm and SVM as a comparison because these two methods in several studies showed good accuracy even though they were considered classical methods [26]–[29]. Fig 8 shows the measurement result between IndoBERT, SVM, and Multinominal Naïve Bayes.

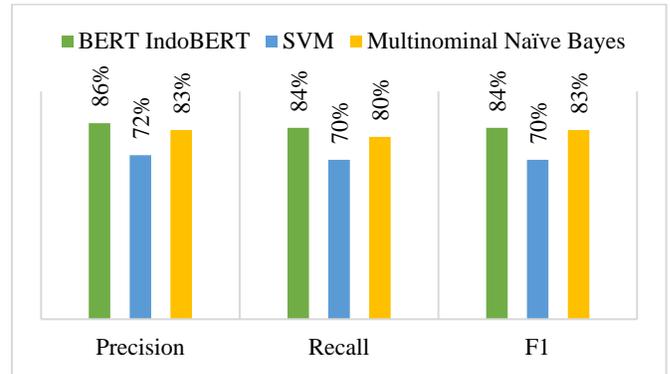

**Fig. 8 Performance Comparison**

Performance Comparison, BERT with IndoBERT model is better performance than SVM and Multinominal Naïve Bayes for the classification model. The higher gap is on the Recall when BERT reaches 84%, Multinominal Naïve Bayes at 80%, and SVM at 70%. As well as precision, BERT reaches the highest recall, 84%, and F1, 84%. This value shows how well BERT carries out the classification process.

## 4. Conclusion

This paper proposes the BERT algorithm with the IndoBERT pre-trained model for sentiment analysis. An experiment with Twitter data on the topic of PPKM Jakarta shows that BERT, especially the IndoBERT model, is better than other algorithms such as SVM and Multinominal Naïve Bayes. In the modern era, understanding social media sentiments is fundamental in seeing people's responses to a topic. BERT is proven to help stakeholders understand the community, especially in making the right policies based on criticism and suggestions from the community.

However, related to huge model training structure and corpus of BERT impact to long execution time. In future research, the execution time issue should be solved by simplifying its model size to improve efficiency. Furthermore, sentiment analysis related to Covid-19 public activity restriction should be explored by visualizing in location mapping. The prediction could be reached using Named Entity Recognition, which possibly extracts location based on tweet text.

**Funding Statement**
Grants from Bina Nusantara University funded this research.